\documentclass[11pt,a4paper]{article}
\usepackage{CJKutf8}
\usepackage[hyperref]{acl2017}
\usepackage{times}
\usepackage{latexsym}
\usepackage{graphicx}
\usepackage{subfigure}
\usepackage{amsmath}
\usepackage{url}
\graphicspath{{Figure/}}
\aclfinalcopy % Uncomment this line for the final submission
\title{Familia: An Open-Source Toolkit for Industrial Topic Modeling}
\author{Di Jiang, Zeyu Chen, Rongzhong Lian, Siqi Bao, Chen Li\\
  Baidu, Inc., China \\
  {\tt \{jiangdi,chenzeyu,lianrongzhong,baosiqi,lichen06\}@baidu.com}   \\}

\date{}

\begin{document}
\maketitle
\begin{abstract}
Familia is an open-source toolkit for pragmatic topic modeling in industry. Familia abstracts the utilities of topic modeling in industry as two paradigms: semantic representation and semantic matching. Efficient implementations of the two paradigms are made publicly available for the first time. Furthermore, we provide off-the-shelf topic models trained on large-scale industrial corpora, including Latent Dirichlet Allocation (LDA), SentenceLDA and Topical Word Embedding (TWE).  We further describe typical applications which are successfully powered by topic modeling, in order to ease the confusions and difficulties of software engineers during topic model selection and utilization.
\end{abstract}

\section{Introduction}

Topic modeling is a well-recognized approach for organizing, searching and understanding the vast amounts of documents \cite{blei2012probabilistic}. In the last decade, various topic models have been proposed in academia. However, only a very small portion of these models have been applied in industry. The major problems that hinder the wide adoption of topic modeling in industry are essentially twofold. First, many research and engineering endeavors are devoted to Probabilistic Latent Semantic Analysis (PLSA) \cite{hofmann1999probabilistic} and its fully Bayesian counterpart Latent Dirichlet Allocation (LDA) \cite{blei2003latent} ,  while the vast majority of the other topics models lack open-source implementation. Second, most of existing research and engineering work focus on ``designing new topic models" rather than ``applying topic modeling results in real-life application''. This discrepancy leads to the fact that there is little work discussing how to properly utilize topic modeling results in real-life scenarios. Given the absence of a comprehensive guide, it is often difficult for developers to select an appropriate topic model for their tasks and to apply the model in a proper way.  

As one step toward bridging the gap between topic modeling research and industrial applications, we open-source a topic modeling toolkit named Familia\footnote{https://github.com/baidu/Familia}. Three off-the-shelf topic models are readily provided in Familia, including LDA, SentenceLDA \cite{jo2011aspect} and TWE \cite{liu2015topical}, which are trained based upon large-scale industrial corpora. Developers have the freedom of exploring more topic models beyond LDA. For effective usage in industrial applications, efficient implementations of two utility paradigms (i.e., semantic representation and semantic matching) have been made publicly available in Familia. We further discuss several industrial cases that benefit from the technique of topic modeling,  aiming to easing the confusions and difficulties for developers during topic model selection and application. The rest of this paper is organized as follows. We detail the open-sourced content of Familia in Section~\ref{sec:Familia Overview}. Then we discuss several industrial cases in Sections  \ref{sec:Industrial Cases}. Finally, we conclude the paper in Section~\ref{sec:Conclusion}.

\section{Familia Overview}
\label{sec:Familia Overview}

Familia is composed of two major contents: source code for two industrial utility paradigms and the trained topic models.

\subsection{Industrial Utility Paradigms}
\label{sec:Utility Paradigms}

The industrial utilities of topic models can be broadly divided into two categories: semantic representation and semantic matching. For semantic representation, Familia provides two kinds of Markov chain Monte Carlo (MCMC) algorithms for users to investigate and choose: Gibbs sampling \cite{griffiths2004finding} and Metropolis-Hastings \cite{yuan2015lightlda}. For semantic matching, Familia contains some functions for calculating the semantic similarity between texts of different lengths: short-long text matching and long-long text matching. Moreover, in order to better display the details of trained topic models, Familia also provides functions such as nearest-word querying, topic-word querying, etc.

\subsection{Trained Topic Models}
Three off-the-shelf topic models of high industrial value are publicly released in Familia. Each of them is trained on the large-scale industrial corpora. The characteristics of the three topics are briefly summarized as follows:
\begin{itemize}
\item LDA: Each document is represented as a mixture over latent topics and each topic is modeled as a distribution of words. 

\item  SentenceLDA: SentenceLDA assumes that the words within one sentence are generated by the same topic. It models the co-occurrence of words in finer granularity than LDA. 

\item TWE:  TWE utilizes LDA topics as the complementary information for training word embeddings. Hence, TWE can provide both topic embeddings and word embeddings.  As the topics derived by LDA are often dominated by words of high-frequency, the embeddings of TWE can partially alleviate this problem by capturing the semantics of low-frequency words under each topic.
\end{itemize}

\section{Industrial Cases}
\label{sec:Industrial Cases}

In this section, we discuss several industrial cases that benefit from the aforementioned paradigms and models. This section acts like a guide for developers to select appropriate topic models for their tasks and apply them in a proper way. 

\subsection{Semantic Representation}

We first discuss some cases involving semantic representation.  The semantic representation derived by topic modeling typically works as features for other machine learning models.

\subsubsection{Document Classification}

\begin{figure*}
\centering
\subfigure[News Topics Distribution as Augmented Features of GBDT]
{
\includegraphics[width=0.3\textwidth]{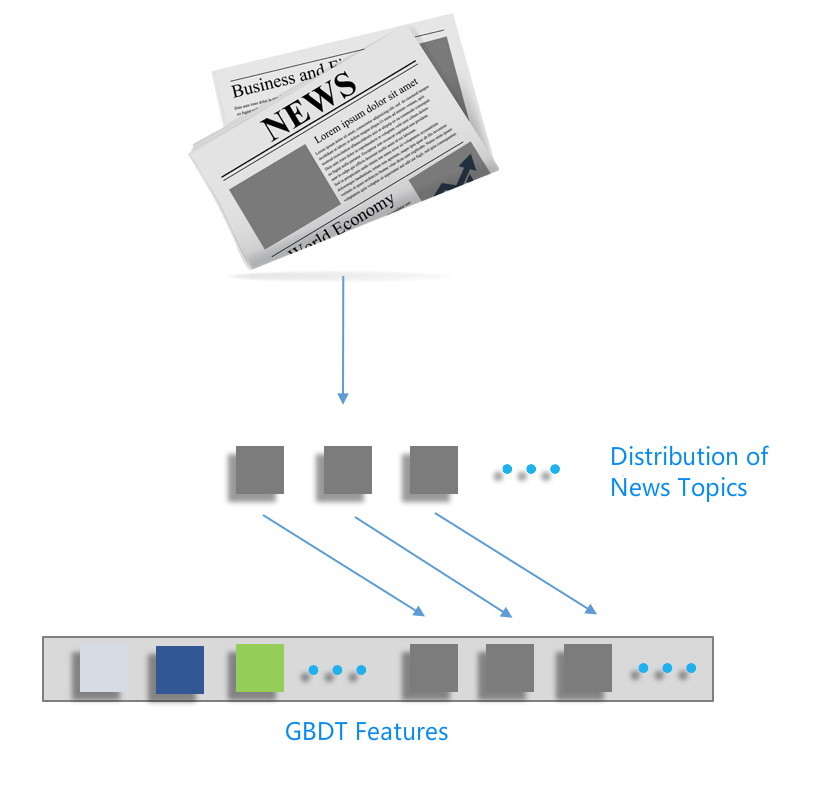}
\label{pic:news_GDBT}
}
\quad
\subfigure[Experimental Results of News Classification] 
{
\includegraphics[width=0.4\textwidth]{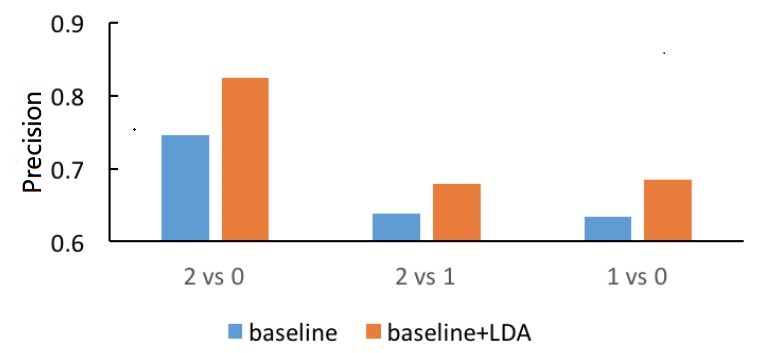}
\label{pic:news_quality}
}
\caption{Classification of News Articles} 
\label{fig:News Classification}
\end{figure*}

The first case is \textbf{classification of news articles}.
For news feed service, the articles collected from various sources often contain low-quality ones. In order to improve user experience, we need to design a classifier to distinguish the good ones from the bad ones. Conventionally, the classifier is built upon some handcrafted features, which include source sites, text length, the total number of images, etc. We could employ topic model to obtain the topical distribution of each article and augment the handcrafted features with this distribution (shown in Figure \ref{pic:news_GDBT}).  As an experiment, we prepare 7,000 news articles, which are manually labeled into 5 categories, in which 0 stands for those of the lowest quality, and 4 represents for the best. We train Gradient Boosting Decision Tree (GBDT) on 5,000 articles with different features and test the trained classifier on the other 2,000 articles.  Figure\ref{pic:news_quality} shows the result from the two classifiers using different sets of features: baseline, baseline+LDA. The results of using features of topic model are significantly better, showing that topic model is an effective way for document representation.

\subsubsection{Document Clustering}
\begin{figure}[h]
\begin{center}
\includegraphics[width=0.5\textwidth]{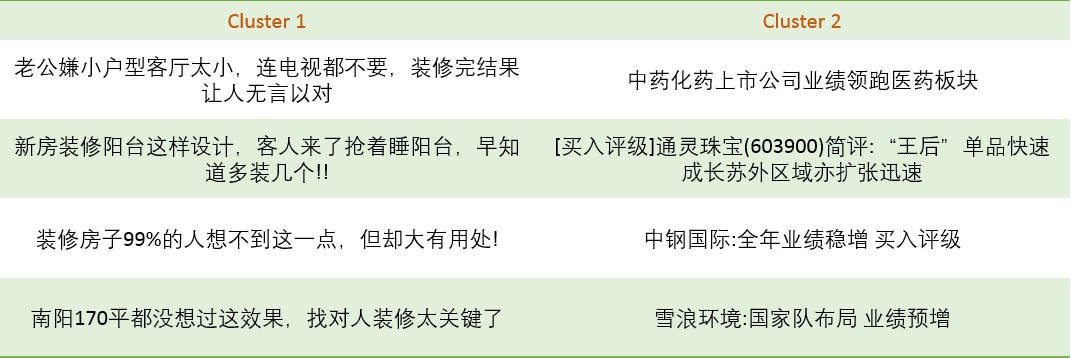}
\caption{Example of Clustering News Articles}
\label{pic:topic model clustering}
\end{center}
\end{figure}

Straightforwardly, the semantic representation of documents could be utilized for clustering. In the task of \textbf{clustering new articles}, we use LDA to compute the topic distribution of news articles and cluster the articles by K-means. Figure~\ref{pic:topic model clustering} shows two clusters which are obtained by clustering 1000 articles into ten groups. Cluster1 is of articles related to interior design and Cluster2 contains articles about the stock market. The result shows that news articles can be semantically clustered based on their topic distributions. 

\subsubsection{Document Information Richness}

In text mining, we frequently encounter the need for evaluating information richness of documents. This requirement can be partially met by topic distributions of documents. We first calculate the topical distribution of the document by topic modeling, and then calculate the entropy of the topical distribution as follows:

\begin{equation}
\label{eq:entropy}
\begin{aligned}
Entropy(d) = -\sum_{i=1}^K{p_i \log p_i},
\end{aligned}
\end{equation}
where $d$ represents document, $K$ represents total number of topics, $p_i$ represents the possibility of the $i$-th topic. The higher the entropy, the higher the richness of document content.  In the scenario of \textbf{web information retrieval},  the richness value is utilized as a feature in sophisticated ranking models.

\subsection{Semantic Matching}

Another paradigm is semantic matching, which can be further categorized as  short-short text matching, short-long text matching and long-long text matching. 

\subsubsection{Short-Short Text Matching}

The need for short-short text matching is common in web search, where we need to compute the semantic similarity between queries and web page titles. Due to the difficulty of topic modeling on short text, embedding-based models such as Word2Vec and TWE are much more common for this task. Assume we want to compute the semantic similarity between a query $q=$``recommend good movies'' and a web page title $t=$``2016 good movies in China'', we first convert the queries into their embeddings (i.e., $\vec{q}$ and $\vec{t}$) and then compute the semantic similarity between these embeddings with the metric of cosine similarity.

\begin{equation}
\label{eq:cos_sim}
\begin{aligned}
CosineSimilarity(\vec{q}, \vec{t}) = \frac{\vec{q} \cdot \vec{t}}{|\vec{q} | |\vec{t}|}
\end{aligned}
\end{equation}

\noindent There are more sophisticated short-short text matching mechanisms in literature, interested readers may refer to deep neural network based models such as Deep Structured Semantic Model (DSSM) \cite{huang2013learning} and Convolutional Latent Semantic Model (CLSM) \cite{shen2014latent}.

\subsubsection{Short-Long Text Matching}

In many online applications, we need to compute the semantic similarity between query and document. Since query is typically short and document content is much longer, short-long text matching is needed in this scenario.  Due to the difficulty of topic inference on short text, we compute the probability of the short text generated from the topic distribution of the long text as follows:

\begin{equation}
\label{eq:cos-sim}
\begin{aligned}
Similarity(q,c)=\prod_{w\in q}\sum_{k}P(w|z_k)P(z_k|c),
\end{aligned}
\end{equation}
where $q$ stands for query, $c$ for document content, $w$ for words in query and $z_k$ for topics.

\begin{figure}[h]
\begin{center}
\includegraphics[width=0.4\textwidth]{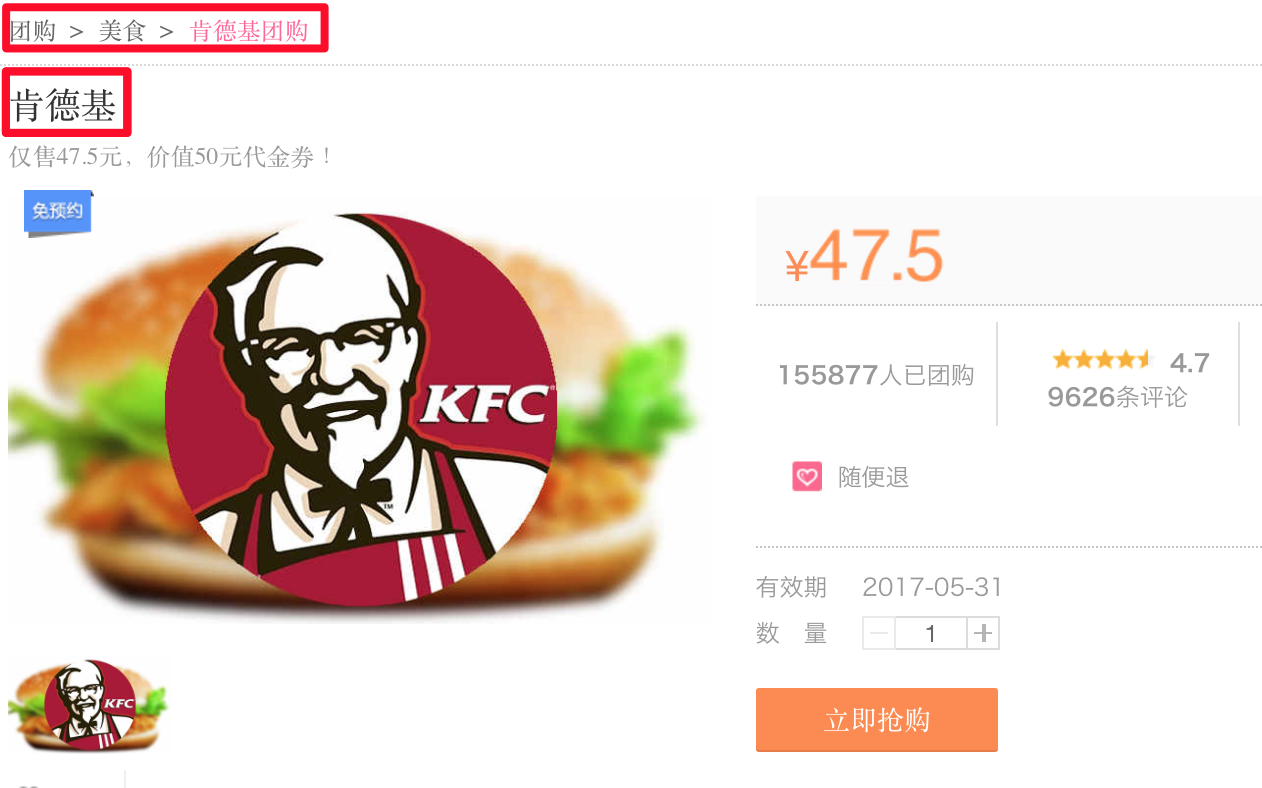}
\caption{Example of Ad Page}
\label{pic:example of nuomi}
\end{center}
\end{figure}
	
\begin{figure}
\centering
\subfigure[Baseline]
{
\includegraphics[width=0.42\textwidth]{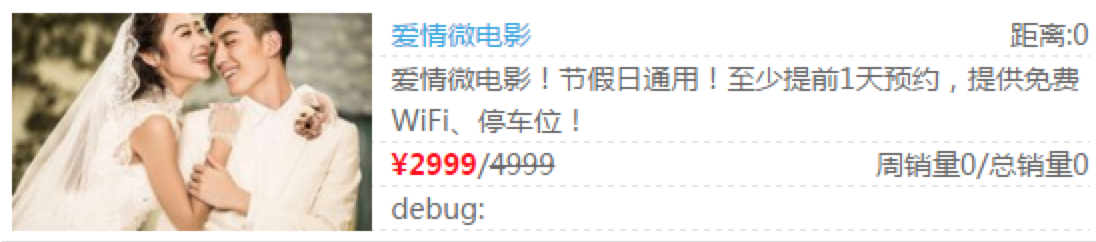}
}
\subfigure[Result with SentenceLDA Feature]
{
\includegraphics[width=0.42\textwidth]{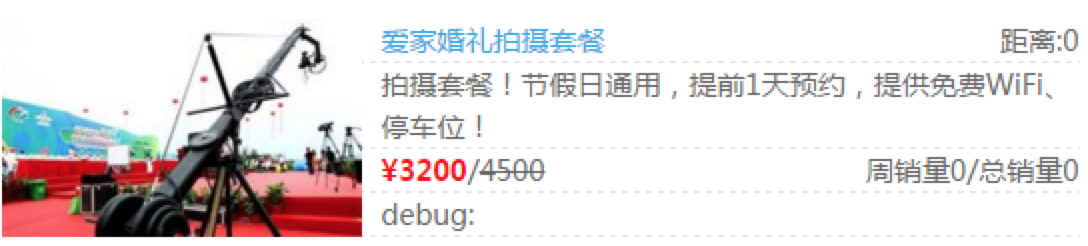}
}
\caption{Semantic Matching of Query-Ad}
\label{fig:user-webpage}
\end{figure}

We first discuss the task of \textbf{online advertising}, in which we need to compute the semantic similarity between query and ad pages. We treat each textual field on ad page as a sentence (shown in Figure \ref{pic:example of nuomi} ) and apply SentenceLDA for this task. After obtaining the topic distribution of each ad page,  we apply Eq.(\ref{eq:cos-sim}) to compute the semantic similarity between query and the ad page. Such similarity can be utilized as a feature in downstream ranking models. For a query ``recording of wedding ceremony'', we compare its ranking results from two strategies in Figure \ref{fig:user-webpage}. We can see that the result with SentenceLDA feature is better at satisfying the underlying need of the query.

\begin{figure}[h]
\begin{center}
\includegraphics[width=0.5\textwidth]{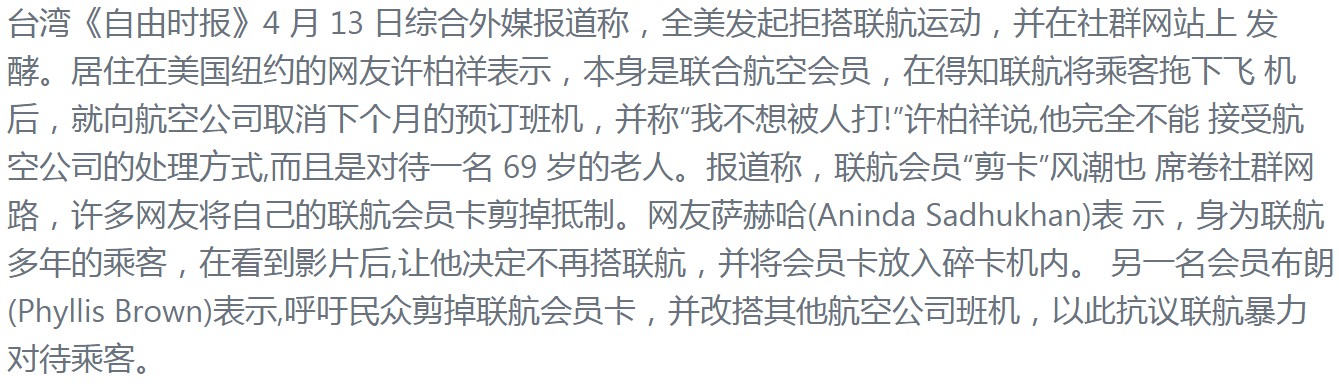}
\caption{Example of News Article}
\label{pic:news-example}
\end{center}
\end{figure}

\begin{figure}[h]
\begin{center}
\includegraphics[width=0.4\textwidth]{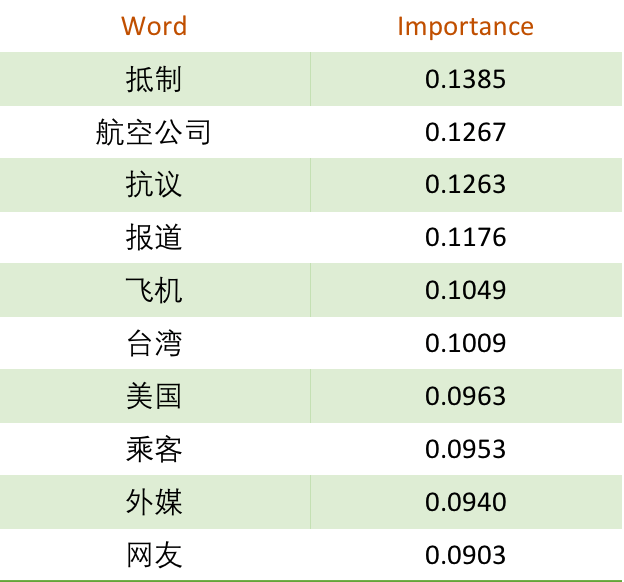}
\caption{Keyword Extraction based on TWE}
\label{pic:keyword}
\end{center}
\end{figure}

An extreme case of short-long text matching is the task of \textbf{keyword extraction from document}.  We extract a set of keywords from documents as concise and explicit representation of the document. The conventional way of extracting keywords from texts relies upon the TF and IDF information. If we want to introduce the semantic importance, we can use Eq.(\ref{eq:TWE keyword}) to compute the similarity of a word and the document as follows:

\begin{equation}
\label{eq:TWE keyword}
\begin{aligned}
Similarity(w, c) = \sum_{k=1}^K{cos(\vec {v_w}, \vec {z_k})P(z_k|c)},
\end{aligned}
\end{equation}
where $c$ stands for document content, $w$ for each word, $\vec{v_w}$ for word embedding for word $w$ and $\vec{z_k}$ for vector representation of topic $z$. Figure~\ref{pic:news-example} is a piece of news article. We use Eq.(\ref{eq:TWE keyword}) to compute the similarity between each word and the whole article. Top-10 keywords (with stop words eliminated) extracted by TWE are shown in Figure \ref{pic:keyword}, and we can see that the keywords from TWE preserve the important information in the news.

%\begin{table}[ht]
%\centering
%\caption{TWE Keyword Extraction}
%\label{table:keyword}
%\begin{tabular}{ | c | c | c | c |}
%\hline
%词语 & 相似度值 \\ \hline 
%抵制 & 0.1385 \\ \hline  
%航空公司 &0.1267 \\ \hline 
%抗议 & 0.1263 \\ \hline 
%报道 & 0.1176 \\ \hline 
%飞机 & 0.1049 \\ \hline 
%台湾 & 0.1009 \\ \hline 
%美国 & 0.0963 \\ \hline 
%乘客 & 0.0953 \\ \hline 
%外媒 & 0.0940 \\ \hline 
%网友 & 0.0903 \\ \hline 
%\end{tabular}
%\end{table}

\subsubsection{Long-Long Text Matching}

We can evaluate the semantic similarity between long texts by the distance of their topical distributions. Such semantic similarity can be further utilized as a feature in various machine learning models.  The distance metrics of gauging two topical distributions include Hellinger Distance (HD) and Jensen-Shannon Divergence (JSD). Hellinger Distance is formally defined as follows:

\begin{equation}
HD(P,Q) = \frac{1}{\sqrt{2}}\sqrt{\sum_{i=1}^{K}(\sqrt{p_{i}}-\sqrt{q_{i}})^{2}},
\end{equation}
where $p_{i}$ and $q_{i}$ are the $i$-th element of the corresponding distributions. The definition of Jensen-Shannon Divergence (JSD) is as follows:
\begin{equation}
\begin{aligned}
JSD(P||Q) = \frac 1 2 (KLD(P||M) + KLD(Q||M))
\end{aligned}
\end{equation}

\begin{equation}
\begin{aligned}
M = \frac 1 2 (P + Q)
\end{aligned}
\end{equation}

\begin{equation}
\begin{aligned}
KLD(P||M) = \sum_{i=1}^K p_i \ln \frac {p_i} {m_i}
\end{aligned}
\end{equation}
where $KLD$ stands for Kullback-Leibler Divergence. 

\begin{figure}[ht]
\begin{center}
\includegraphics[width=0.4\textwidth]{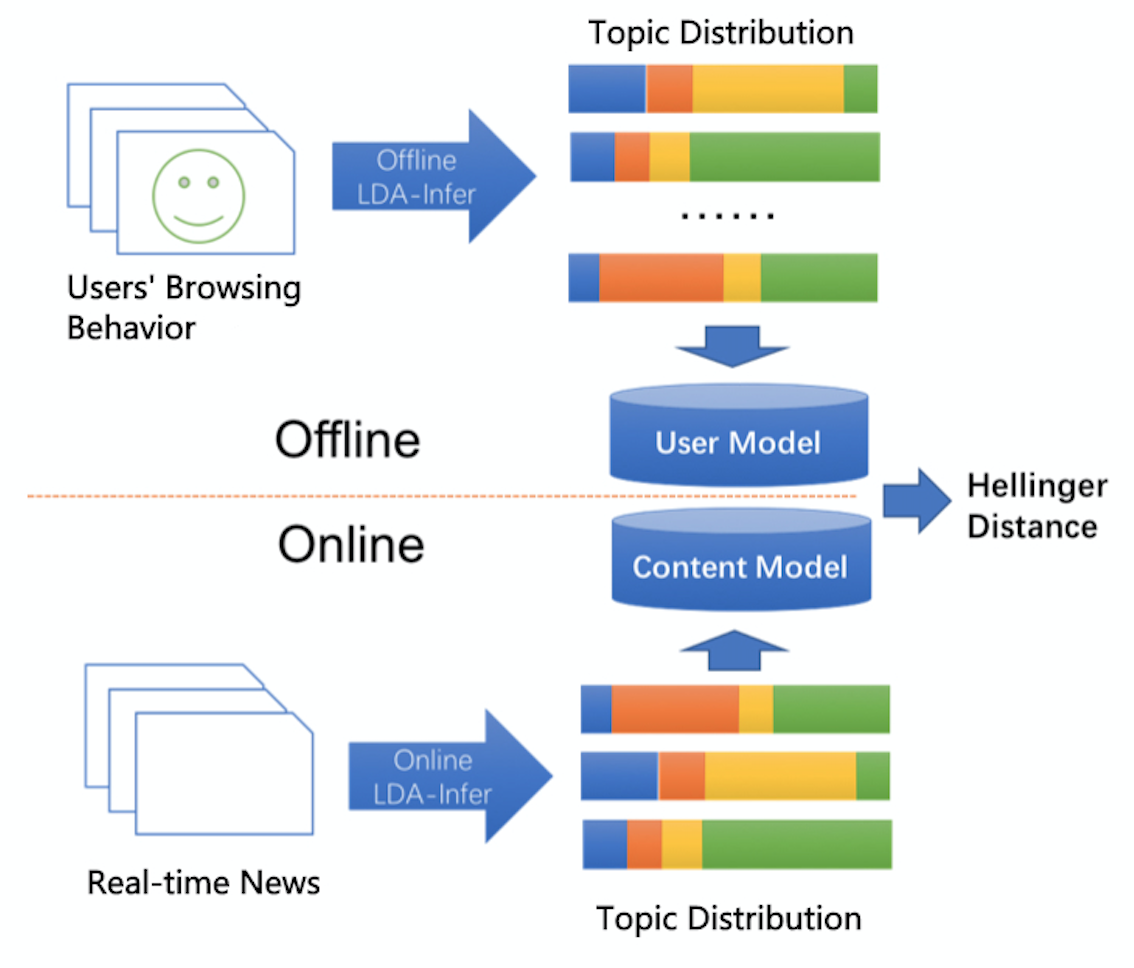}
\caption{Personalized News Recommendation}
\label{pic:news recommendation}
\end{center}
\end{figure}

We now discuss the task of \textbf{personalized news recommendation}, which is illustrated in Figure \ref{pic:news recommendation}. We first collect the news articles (or news titles) recently read by each user and compose them into a pseudo document. By conducting topic modeling on a corpus of pseudo documents, we obtain topic distribution of each pseudo document, and this distribution works as the corresponding user profile. In online setting, we compute the Hellinger Distance(HD) between the topic distribution of real-time news articles and the user profile, and those with low HD value are pushed to the user as personalized news feed. This approach has effectively improved the performance of real-world news feed application.

\begin{figure}[h]
\begin{center}
\includegraphics[width=0.5\textwidth]{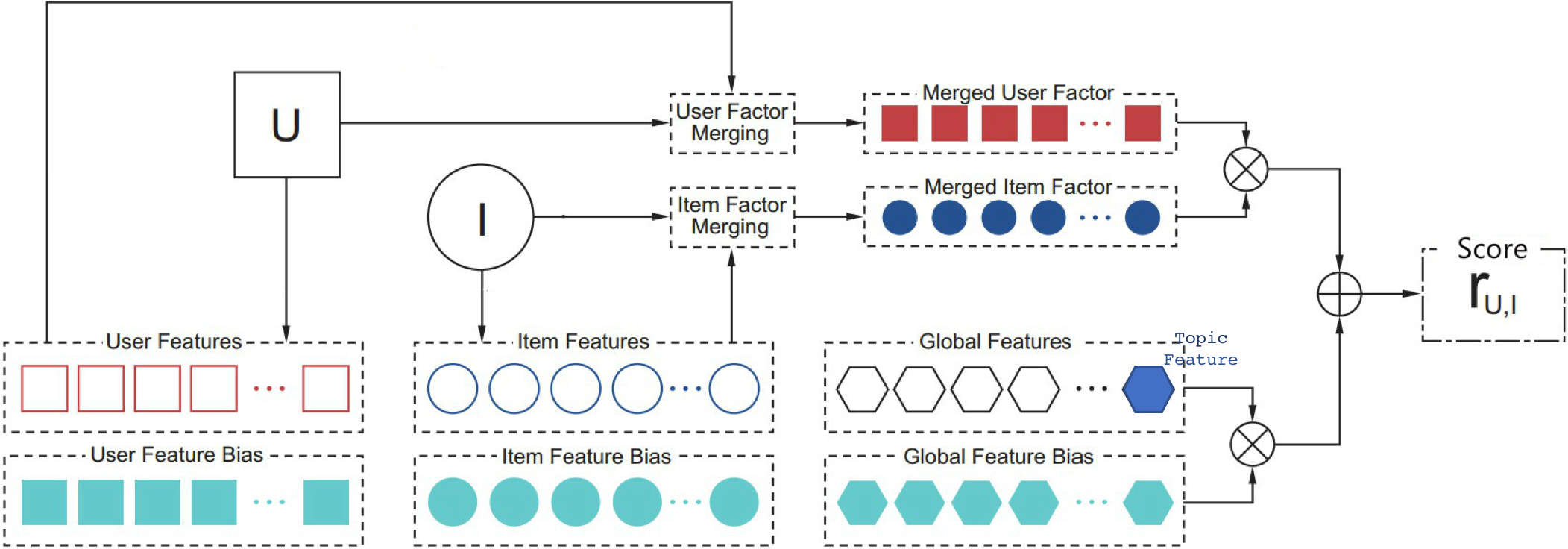}
\caption{SVDFeature with Topic Feature}
\label{pic:feature-based}
\end{center}
\end{figure}

We proceed to discuss the task of \textbf{personalized fiction recommendation}. Matrix factorization is a common approach for industrial recommendation systems. SVDFeature \cite{chen2012svdfeature} is a framework designed to efficiently solve the feature-based matrix factorization. SVDFeature is quite flexible and is able to accommodates global features, user features and item features. SVDFearure can be mathematically described as follows:

\begin{equation}
\small
\begin{aligned}
\label{eq:svdfeature}
y = \mu + (\sum_{j}b_{j}^{(g)}\gamma_{j} + \sum_{j}b_{j}^{(u)}\alpha_{j} + \sum_{j}b_{j}^{(i)}\beta_{j}) + \\
(\sum_{j}p_{j}\alpha_{j})^{T}(\sum_{j}q_{j}\beta_{j}),
\end{aligned}
\end{equation}
where $y$ is target, $\mu$ is a constant indicating the global mean value of target, $\alpha$ represents user feature, $\beta$ represents item feature, $\gamma$ represents global feature, $b^{(g)}$ is weight of global feature, $b^{(u)}$ is weight of user feature, $b^{(i)}$ is weight of item feature, $p$ and $q$ are model parameters.

\begin{figure}[h]
\begin{center}
\includegraphics[width=0.4\textwidth]{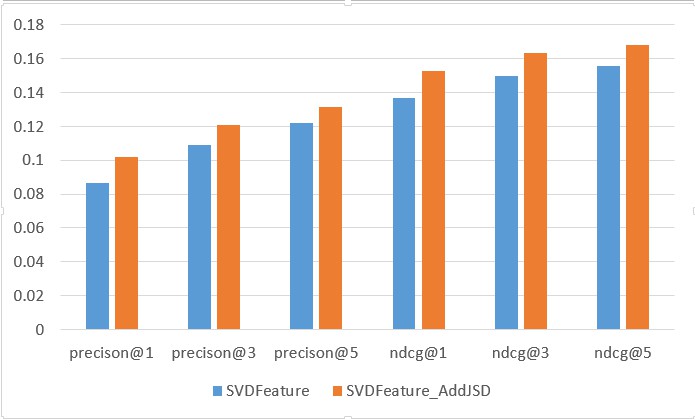}
\caption{Fiction Recommendation Performance}
\label{pic:Fiction Recommendation Performance}
\end{center}
\end{figure}

In the scenario of personalized fiction recommendation, each user has some historically downloaded fictions. By conduct topic modeling on these fictions, we can obtain the user's topic representation, which works as a user profile of reading interests. By computing the JSD between the topic distribution of each fiction and the user profile, we can quantify the probability that user is interested in this fiction. We augment the aforementioned SVDFeature framework with the JSD value as a global feature (Figure \ref{pic:feature-based}). From a comparative study shown in Figure~\ref{pic:Fiction Recommendation Performance}, we can see that adding JSD is effective to improve the performance of SVDFeature. SVDFeature with JSD constantly outperform its original counterpart in terms of both Precision and NDCG.

\section{Conclusion}
\label{sec:Conclusion}

Familia is an open-source toolkit for industrial topic modeling. It supports two industrial utility paradigms: semantic representation and semantic matching.  Three topic models of high industrial value have been made publicly available as well. These topic models are all trained based upon large-scale industrial corpora across various domains. We present several industrial cases where the technique of topic modeling is successfully applied.  The discussion of these cases works as a guide for developers to conduct topic model selection and utilization in their own tasks.  We wish Familia could help engineers to employ the technique of topic modeling in more convenient way and inspire more pragmatic research on topic models.

\bibliography{acl2017}
\bibliographystyle{acl_natbib}

\end{document}